\documentclass{WileyMSP-template}
\usepackage{hyperref}
\usepackage{xcolor}

\usepackage{ragged2e}

\justifying

\begin{document}
\pagestyle{fancy}
\rhead{\includegraphics[width=2.5cm]{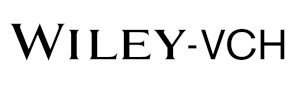}}

\title{Fundamental Phase Noise in Thin Film Lithium Niobate Resonators}
\maketitle

\author{Ran\ Yin$^{1,2,\dagger}$}
\author{Yue\ Yu$^{1,2,3,\dagger}$}
\author{Chunho\ Lee$^{1,2}$}
\author{Ian\ Christen$^{1}$}
\author{Zaijun\ Chen$^{1,2}$}
\author{Mengjie\ Yu$^{1,2,3,*}$}

\begin{affiliations}
\noindent1 Department of Electrical Engineering and Computer Sciences, University of California, Berkeley, California 94720, USA \\
2 Ming Hsieh Department of Electrical and Computer Engineering, University of Southern California, Los Angeles, CA, 90089, USA \\
3 Materials Sciences Division, Lawrence Berkeley National Laboratory, Berkeley, CA, USA \\
$\dagger$ The authors contribute equally\\
$*$ Email Address: mengjie.yu@berkeley.edu
\end{affiliations}


\begin{abstract}
Fundamental phase noise in thin-film lithium niobate (TFLN) photonic integrated circuits is governed by thermal-charge-carrier-refractive (TCCR) dynamics arising from thermally driven carrier fluctuations. In contrast to the predominantly thermorefractive noise in silicon photonic platforms, TCCR noise represents a distinct mechanism that becomes critical for applications requiring high frequency stability and phase coherence, including optomechanical sensing, low-phase-noise microwave synthesis, and on-chip quantum squeezing.
A quantitative understanding of the deterministic parameters that control TCCR noise is therefore essential for engineering the next generation of low-noise TFLN photonic systems. 
Here, we identify two dominant contributors to the TCCR noise in TFLN microresonators: material anisotropy and surface states. Material anisotropy results in increased noise for extraordinarily polarized optical modes and leads to a geometry dependent phase noise. Surface-state effects manifest as increased noise in higher-order transverse modes as well as more than 120-fold higher noise in suspended microresonators. Finally, we demonstrate that post-fabrication annealing --- widely used to reduce defect densities and recover crystal quality --- suppresses frequency noise by a factor of 8.2 in cladded microresonators. Together, these results establish a practical pathway for noise engineering in TFLN integrated photonic devices and accelerate their deployment in next-generation precision photonic systems. © 2025 The Author(s)

\end{abstract}

\section{Introduction}
$\,\,\,$ Photonic integrated circuits (PICs) offer a scalable and energy-efficient solution for realizing functions that traditionally require bulky optical systems. Among these, microresonators --- combining ultra-high quality factors ($Q$) with small mode volumes --- have enabled diverse applications, including optomechanical sensing \cite{LeiLiu2021_optomech_sens, Guo2022_quant_optomech, optomech_measure_precision}, low-noise microwave synthesis \cite{Lei2022, Cheng25}, narrow-linewidth lasers \cite{Zhao2024,Liang2015, alk2023}, and quantum photonic computing and sensing \cite{15dB_squeeze_application, Wang2025, quantum_compt_mode, Shaoyuan, Paraïso2021_quantum_computing}. The small optical mode volume, which contributes to the enhanced Purcell factor and compact form factor of the devices, also leads to stronger refractive-index fluctuations arising from fundamental thermodynamic and charge-induced noise, in particular in materials which possess complex properties or high defect densities. As a result, index fluctuations set the fundamental noise floor in any on-chip coherent photonic system which is sensitive to or relies on optical phase stability or detection, thereby constraining application-specific performance metrics such as dynamic range in optomechanical measurements, phase noises in laser and microwave synthesis, and quantum advantages in continuous-variable-based photonic computing and sensing.

\begin{figure}[h]
  \centering
  \includegraphics[width=0.8\linewidth]{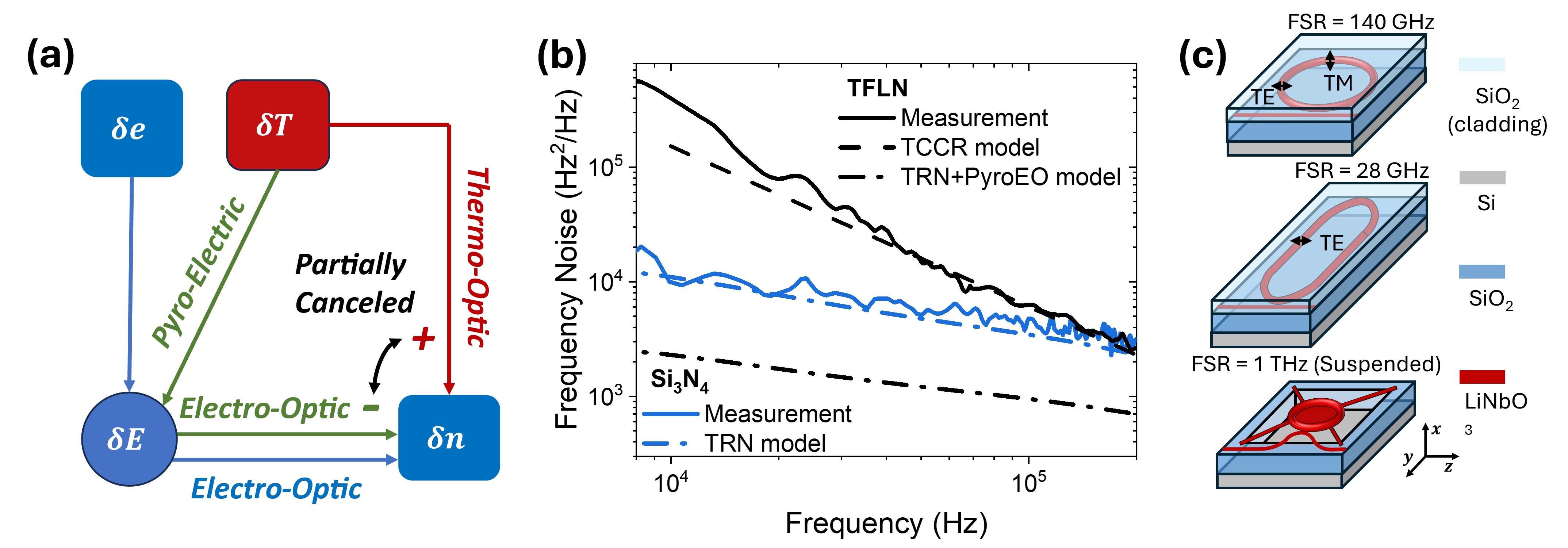}
  \caption{
  (a) Sources of the fundamental noise for TFLN microresonators. 
  (b) Fundamental noise of microresonators on Si$_3$N$_4$ and TFLN platform, respectively. The free spectral range of the X-cut TFLN ring resonator and Si$_3$N$_4$ ring resonator are 140 GHz and 100 GHz, respectively. The input light is TE polarized.
  (c) Microresonators samples that are studied in this work.
  }
  \label{fig:boat1}
\end{figure}

Fundamental phase and frequency noise has been investigated on various chip-based platforms that support high-$Q$ microresonators, including photonic crystal cavities on silicon \cite{sio_noise}, microspheres on silica \cite{Gorodetsky:04}, ring resonators on silicon nitride (Si$_3$N$_4$) and tantalum pentoxide (Ti$_2$O$_5$) \cite{SiN_noise, Liu2025_Tantalum_Pentoxide}, and whispering gallery resonators on magnesium fluoride \cite{Lim2017_MgF}. In these centrosymmetric materials, the frequency noise is primarily dominated by thermorefractive noise (TRN), where stochastic temperature fluctuations modulate the refractive index through the thermo-optic effect. TFLN, benefiting from its diverse material properties like ferroelectricity, pyroelectricity, optical nonlinearities, and a strong electro-optic (EO) effect, has opened versatile opportunities for integrated photonics \cite{Ren2025} \cite{LN_application}. However, these rich intrinsic properties also introduce additional noise contributions to LN that is distinct from centrosymmetric materials. 
In TFLN, the thermal noise ($\delta T$) is coupled to the refractive noise ($\delta n$) through through two pathways (Figure~\ref{fig:boat1}a): the positive thermo-optic effect (thermal refractive noise, TRN), and the joint combination of the pyroelectric effect and EO effect (pyro-EO noise) which is negatively related to the temperature variation \cite{Ren2025} \cite{Zhang2025}. A recent study reports an additional and dominant thermal-charge-carrier-refractive (TCCR) noise in TFLN platform, arising from the thermally induced Brownian motion of charge carriers ($\delta e$). This same microscopic mechanism underlies Johnson–Nyquist voltage noise in the electrical domain and, in the optical domain, generates electric-field fluctuations that modulate the refractive index \cite{Zhang2025}. 
In this work, we investigate refractive-index noise in Si$_3$N$_4$ and X-cut TFLN microring resonators. The TFLN microresonator has a free spectral range (FSR) of 140 GHz, with a waveguide width of 1.2 $\mu$m and a height of 0.6 $\mu$m, while the Si$_3$N$_4$ resonator has an FSR of 100 GHz, a width of 1.6 $\mu$m, and a height of 0.75 $\mu$m. Both devices exhibit high loaded $Q$ --- $4.2\times10^6$ for TFLN and $2\times10^6$ for Si$_3$N$_4$ --- enabling sensitive measurements of intrinsic refractive-index noise and access to the fundamental noise limits of each platform. We observe that TCCR noise in TFLN resonators exceeds the thermorefractive noise measured in Si$_3$N$_4$ resonators in the low frequency region (below 200 kHz). Additionally, TCCR noise exhibiting a flicker-like spectrum, with $S_v \propto f^{-1.4}$ scaling, deviating significantly from the TRN frequency scaling observed in Si$_3$N$_4$ resonators (Figure \ref{fig:boat1}b).

Identifying the dominant mechanisms underlying the TCCR noise and developing practical mitigation strategies are critically important to building future TFLN photonic circuits. Here, we systematically investigate the mechanism governing the TCCR noise in TFLN by experimentally characterizing microresonators with varied geometries and optical polarizations (Figure \ref{fig:boat1}c). We find that optical modes polarized along the Z-axis (the polar axis) exhibit higher TCCR noise due to the stronger EO coefficient. Device geometry of X-cut TFLN microresonators (racetrack or ring) also affects the noise intensity due to the different proportion of exposure to the polar axis. Furthermore, the surface-defect density associated with cladding materials and the fabrication process plays a critical role for the charge noise, resulting in over two orders of magnitude higher noise for air-cladded (suspended) microresonators compared to silicon-dioxide-cladded ones. Moreover, we measured elevated noise in higher-order transverse mode as compared to the fundamental mode, indicative of the stronger index coupling to the surface states. Finally, we discover that post-fabrication thermal annealing, which restores the crystalline quality of LN, suppresses the TCCR noise by a factor of 8.2.

\section{Simulation of the TRN, Pyro-EO, and TCCR noise}
$\,\,\,$ We adopt the fluctuation–dissipation model to simulate resonance-frequency noise induced by temperature fluctuations \cite{ Zhang2025, levin98_LIGO1, LEVIN20081941}. In this model, a heat source with a spatial distribution matched to the optical mode profile is applied, and the resulting heat-dissipation rate of the microresonator with varying source modulation frequencies is then used to obtain the corresponding TRN spectrum.
Additionally, we incorporate a net thermo-optic coefficient in the model to account for the coherent cancellation between TRN and pyro–electro-optic noise.

The TCCR noise, arising from the thermodynamic charge fluctuations, can be explained using a simplified model that treats the microresonator as an effective resistor-capacitor system, which is expressed as \cite{Zhang2025}:

\medskip
\begin{equation}
S_{\nu,TCCR}(f, T)=\frac{n^4r^2\nu^2k_BT}{4\pi^2V_{eff}} \frac{\sigma(f, T)}{\epsilon_0^2\epsilon_r^2f^2}
\label{eq:noise}
\end{equation}
\medskip

where $k_B$ is the Boltzmann constant, $v$ is the optical frequency, $V_{eff}$ is the effective optical mode volume, and $\sigma(f, T)$ is the frequency- and temperature-dependent electrical conductivity. Equation~\ref{eq:noise} predicts that the TCCR noise is strongly associated with the anisotropic properties of LN (through $r$), the effective optical mode volume $V_{eff}$, and the conductivity of the microresonators $\sigma(\omega, T)$ which is influenced by surface states. The dependence of the resonance-frequency noise on these parameters is experimentally investigated and compared with the models. 
Material constants and structural parameters employed in the simulations are summarized in the Supplementary Information. 

\section{Experimental setup and results}
\begin{figure}[h]
  \centering
  \includegraphics[width=0.8\linewidth]{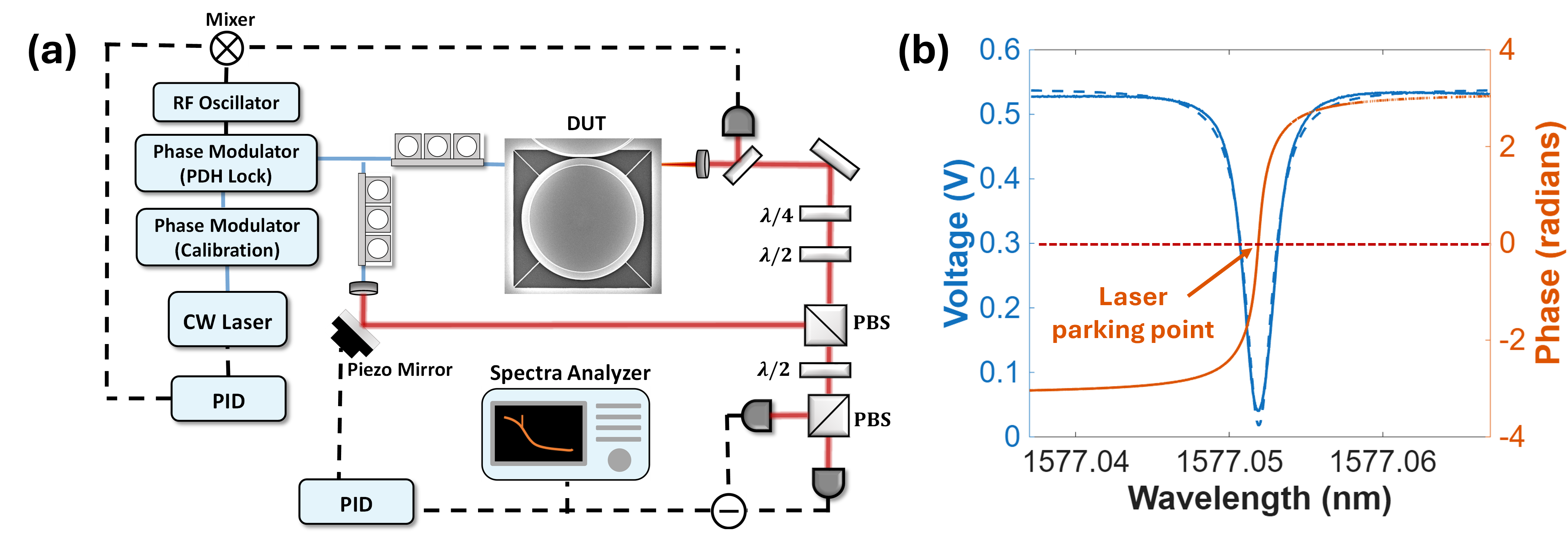}
  \caption{a) Simplified diagram for the noise measurement setup. b) Intensity and phase transmission diagram. The red arrow indicates the laser parking point where the frequency noise is measured. CW: continuous wave; PID: Proportional–Integral–Derivative control system. DUT: device under test. PBS: polarizing beam splitter.} 
  \label{fig:boat2}
\end{figure}

$\,\,\,$ We employ a balanced homodyne detection to measure the fundamental noise spectra of the microresonators (Figure \ref{fig:boat2}a). 
A Pound–Drever–Hall (PDH) locking system is used to stabilize a laser at the center of the resonance dip, where the phase transmission spectrum has the maximum slope (Figure \ref{fig:boat2}b), enabling efficient transduction of the resonance frequency noise into the optical phase noise. This phase noise is then converted to amplitude modulation via interference with a strong reference local oscillator (LO) beam and detected by a balanced photodetector to suppress the laser intensity noise. In addition, a 1-MHz phase modulation signal with a known modulation index is applied for calibration of the resonator frequency noise. Details of the procedure to obtain the resonance frequency noise spectrum are provided in the Supplementary Information.
 
\subsection{Phase noise versus nonlinear anisotropy}
$\,\,\,$ We first examine how the nonlinear anisotropy of LN influences the fundamental noise by measuring under different optical polarizations. Figure \ref{fig:boat3}a shows that an X-cut ring  with a 140-GHz FSR measured with a transverse-electric (TE)--polarized light exhibits approximately 4 times higher noise than it with a transverse-magnetic (TM)--polarized light, which results from the EO coefficient $r$ difference in the TCCR model (Equation~\ref{eq:noise}). For this model, we consider only charge fluctuations along $z$-axis due to the correspondingly strongest EO coefficients. We adopt $r_{TE} = (r_{13}+r_{33})/2 = 20.5$ pm/V as the effective EO coefficient for the TE-polarized light since the circulating optical mode continuously samples different crystallographic orientations, leading to an effective nonlinear response that reflects a geometry-dependent averaging of the anisotropic nonlinear tensor.  The EO coefficient of $r_{TM} = r_{13} = 10$ pm/V is employed to the TM-polarized light. Since TCCR noise is proportional to $r^2$, the simulated noise matches well with the fourfold difference between the frequency noises of TE- and TM- polarized light, as shown in Figure \ref{fig:boat3}a. In addition, we observe that the TCCR noise dominates the total frequency noise for the TE-polarized light below 200 kHz offset frequency (our measurement window) while for the TM-polarized mode the measured noise starts to deviate from the TCCR noise model starting at around 100 kHz. This also agrees well with our simulation when taking the TRN+PyroEO noise source into account where the TRN+PyroEO noises become comparable to the TCCR noise at frequencies above 100 kHz (Figure \ref{fig:boat3}a).

\subsection{Phase noise versus resonator geometry}
\begin{figure}[h]
  \centering
  \includegraphics[width=0.8\linewidth]{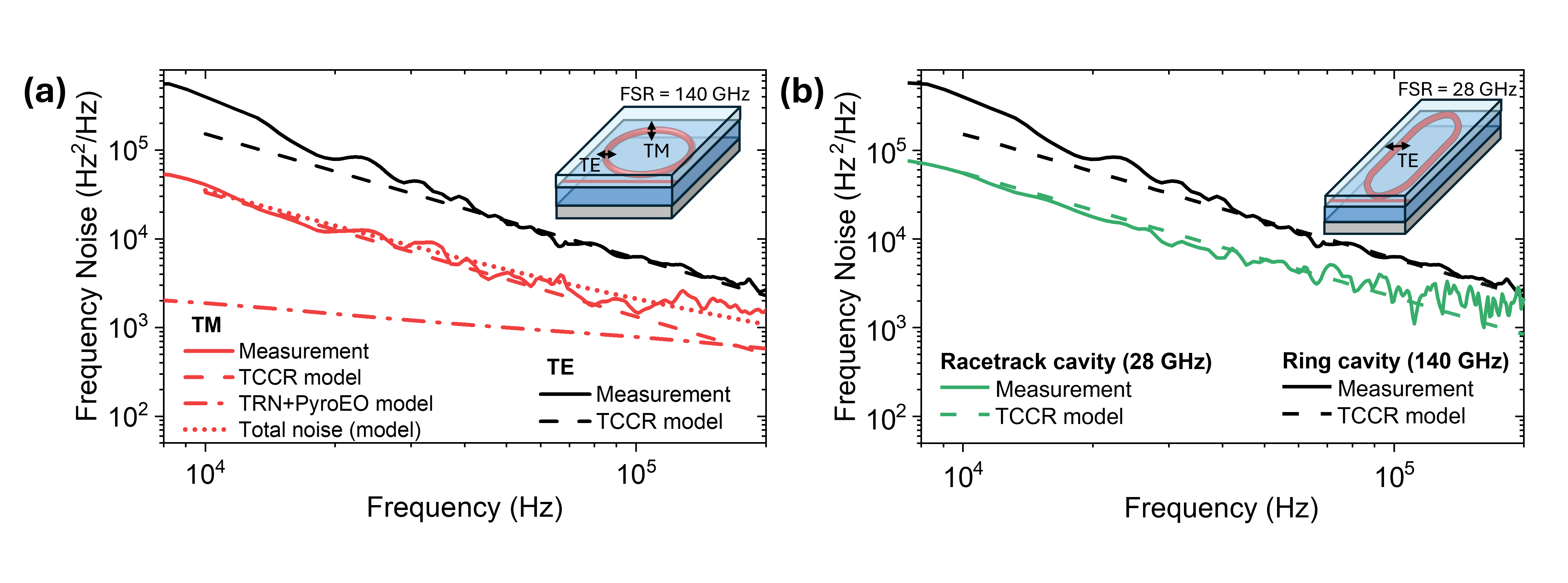}
  \caption{The nonlinear anisotropy of LN affects the fundamental noise of the microresonator, which is demonstrated by single-sided power spectral density (PSD) of frequency noise for: 
  (a) an X-cut LN ring-based microresonator under the interrogation of the TE and TM light polarization and
  (b) an x-cut LN ring-shaped resonator (140-GHz FSR) and racetrack-shaped resonator (28-GHz FSR) under TE light polarization. 
  Experimental results are plotted using solid curves while the simulation result of the TCCR and the summed TRN+pyroEO noise are shown in dashed and dashed dot curves, respectively. The noise in the TM-polarized case has comparable contributions from both the TRN+PyroEO noise and the TCCR noise at frequencies higher than 100 kHz, while the noise in the TE case is primarly dominant by the TCCR noise term. The EO coefficients used for the TE and TM mode in ring-shaped resonator are $r_{TE} = (r_{13}+r_{33})/2$ and $r_{TM} = r_{13}$, respectively, while the EO coefficient for the TE mode in racetrack-shaped resonator is $r_{33}$.}
  \label{fig:boat3}
\end{figure}

$\,\,\,$ The overlap between the light polarization and the polar axis in X-cut TFLN platforms could be utilized when designing the optimal microresonator geometry for different applications. For example, racktrack microresonators with different orientation in plane could be used for achieving strongest EO modulation in broadband EO frequency comb generation \cite{Mian2019_EOcomb}, or to suppress Raman oscillation for facilitating Kerr microcomb generation \cite{MJ_raman_supress, Qifan_Raman_2025, loncar_raman_kerrcomb_2025, rebecca_2024_raman_comb}. Here we demonstrate the geometric difference in resonators also leads to different frequency noise. We compare the frequency noise for the TE-polarized light in a racetrack resonator with a 28-GHz FSR and a ring resonator with a 140-GHz FSR in Figure \ref{fig:boat3}b. 
On one hand, the larger optical mode volume in the racetrack resonator contributes to lower noise (Equation \ref{eq:noise}). 
However, a competing effect arises from that the long straight waveguide sections on the racetrack along the Y-axis direction increases the effective EO coefficient closer to $r_{33}$. Considering the net effects of both mode volume and geometry, a good agreement between measurement and simulation is achieved in Figure \ref{fig:boat3}b.
Additionally, we predict that rotating the racetrack by 90 degrees in plane could have further reduced the TCCR noise by a factor of 9 due to a weaker effective EO coefficient. 

\subsection{Phase noise versus surface state}
\begin{figure}[h]
\centering
\includegraphics[width=0.8\linewidth]{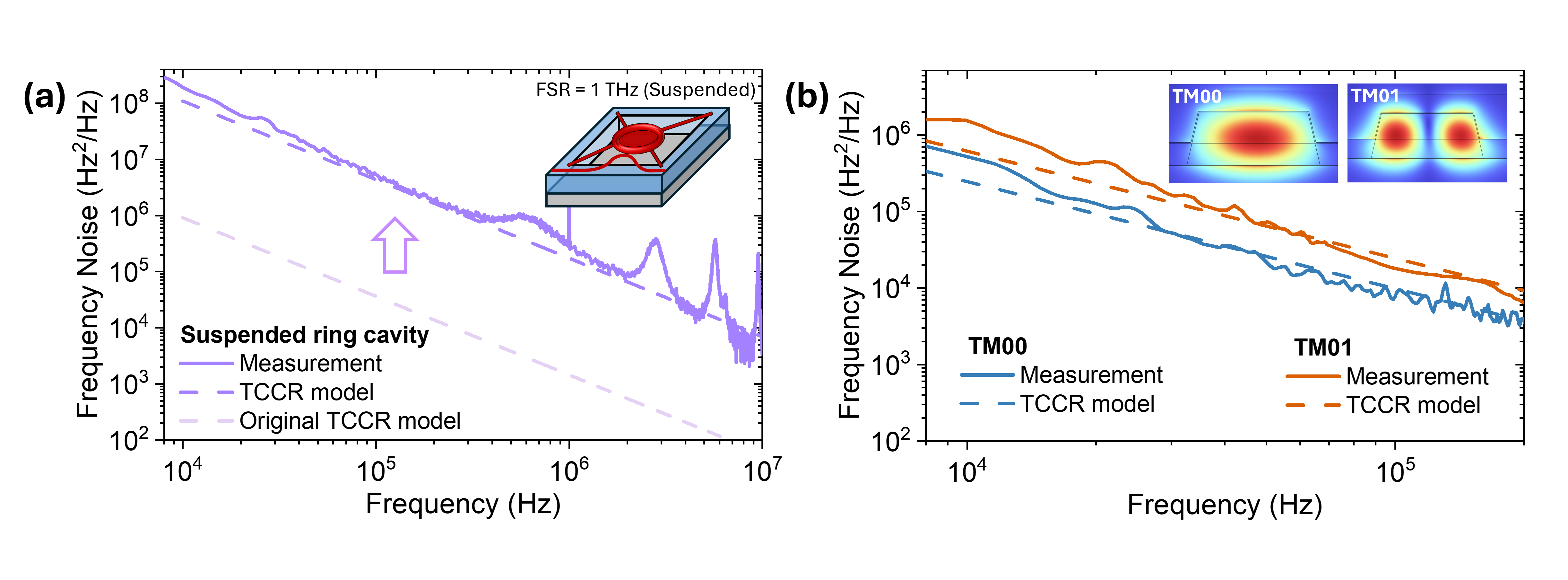}
\caption{The surface states of LN microresonators affect the fundamental noise, which is demonstrated by single-sided power spectral density (PSD) of frequency noise for:
(a) an X-cut LN suspended ring-shaped resonator (air-cladded on both surfaces); The device is fabricated on a 300-nm-thick wafer with a 250-nm etch depth and suspended via a wet etch process. The TCCR model for the suspended device uses an effective conductivity of $\sigma = 4.32\times10^{-7},\mathrm{S/m}$ at a 1-kHz offset frequency, 120 times larger than the conductivity used in the original TCCR model. The peaks observed between 1 and 10 MHz correspond to mechanical modes supported by the suspended structure.
(b) an X-cut LN cladded ring-shaped microresonators measured at TM00 and TM01 optical modes. The device is fabricated on a 600-nm-thick wafer with a 350-nm etch depth and cladded with PECVD silicon dioxide of 800 nm.}
\label{fig:boat4}
\end{figure}

$\,\,\,$ The conductivity $\sigma$ governing the charge noise properties is sensitive to the surface states of the devices. We measure the noise of suspended microresonators, i.e. with all surfaces air-cladded (Figure \ref{fig:boat4}a). The suspended microresonator shows around 120 times higher noise than our model predicts for the cladded case. The trend for the measured noise still follows a $f^{-1.4}$ dependence, indicating that the charge noise is still the dominant noise source, but with a higher conductivity $\sigma$. We attribute this to a higher surface conductivity and a stronger surface contribution due to the increased air-LN interfaces in the suspended resonator. In addition, the suspended resonator is fabricated on a 300-nm-thick LN wafer with a 250-nm etch depth, leading to an increased surface-to-bulk ratio as compared to the other 600-nm-thick TFLN devices reported in the work. To account for the surface effects, the conductivity used to model the suspended device is $4.32\times 10^{-7} S/m$ at 1 kHz offset frequency, compared to the conductivity value of $3.6\times 10^{-9} S/m$ adopted in the original TCCR model.
The effect of the surface states are also recently reported to lead to losses in TFLN based acoustic cavities \cite{surface_bulk_TLS_acoustic}. We observe various mechanical modes of the suspended devices via optomechanical coupling at 1-10 MHz range (Figure \ref{fig:boat4}a). The dynamic range or signal to noise ratio of the optomechanical signals is limited by the increased charge noise rather than the conventional thermo-refractive noise.

The surface states can also explain the observed noise dependence on the transverse optical modes with different spacial field distributions (Figure \ref{fig:boat4}b). We discover that despite the TM01 mode with a larger mode volume, it exhibits 2.3 times higher noise than the TM00 mode, which we attribute to a larger fraction of the TM01 optical field near the surface, where the higher defect density introduces additional charge-fluctuation sources and thus stronger localized electric-field fluctuations that drive refractive-index noise. The role of the surface state is expressed as the increase of the conductivity in the model since the surface defects serve as an additional charge carrier source. To account for the stronger interaction of TM01 mode with surface states, the averaged conductivity is increased to be 2.3 times which matches well with the measurements.

\subsection{Mitigating noise via thermal annealing}
\begin{figure}[htbp]
  \centering
  \includegraphics[width=0.8\linewidth]{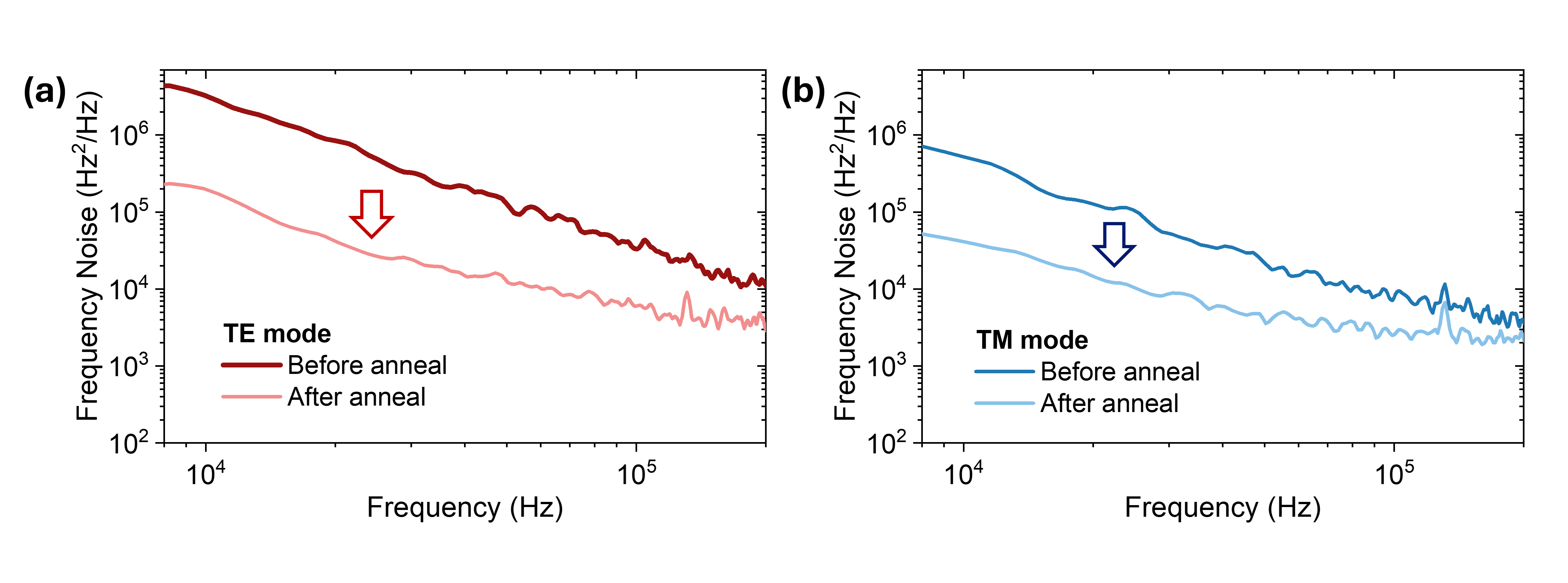}
  \caption{Decreasing the fundamental noise of LN microresonators via annealing, demonstrated by single-sided power spectral density (PSD) of frequency noise for an X-cut LN ring-based microresonator measured with TE polarized mode before and after annealing  
  (a) with a TE polarized optical mode;
  (b) with a TM polarized optical mode.
  }
  \label{fig:boat5}
\end{figure}

$\,\,\,$ Finally, we demonstrate that we can further mitigate the TCCR noise by post fabrication via thermal annealing. 
Post-fabrication annealing has been employed in CMOS devices to effectively decrease flicker noise through lowering the defects density that serve as charge traps in electrical domain \cite{Mosfet_anneal_1}.  For this study, we anneal the TFLN chips at 500 degree for 2 hours under atmosphere. Since the reduction of the conductivity by annealing is intrinsic to the material itself, annealing shows a polarization independent behavior (both polarizations are sensitive to $E_z$) and lowers the frequency noise level by a factor of 8.2 for both TE- and TM- polarized light, as shown in Figure~\ref{fig:boat5}. To note, the previously reported suspended devices can not be annealed thermally to restore the surface quality due to its vulnerability of thermomechanical expansion leading to device collapses. 

\medskip

\section{Discussion and conclusion}
$\,\,\,$ In conclusion, we demonstrate that the phase and frequency instability on the TFLN platform is primarily determined by the charge fluctuation and higher as compared to the Si$_3$N$_4$ platform, in particular at low frequency range (below 200 kHz for oxide-cladded devices and below 10 MHz for suspended air-cladded devices). Phase noise originates from the strong anisotropic nonlinearities of the material and is sensitive to the surface conditions and the fabrication process, as evidenced by noise measurement when probing different spatial modes and devices with different cladding materials. We observe a conductivity increase of more than two orders of magnitude in devices fabricated on a thinner TFLN wafer and with more LN interfaces exposed to air.  The increased noise floor is evidenced by a reduced signal-to-noise ratio of observed optomechanical signals spanning 1 to 10 MHz, and must be explicitly considered for designing the optomechanical sensors for ultrasound imaging, mass and weak force detection. The findings strongly indicate that the thin film material behaves distinctively different from its bulk form, depending on the surface-area-to-volume ratio and charge dynamics associated with crystallographic anisotropy.  The experimental results show excellent agreement with simulations that incorporate the optical, electrical, and thermal material properties as well as the detailed device geometry. Although this study focuses on X-cut TFLN wafers, the analysis framework is general and can be extended to other crystal orientations by adopting the appropriate material tensors. In X-cut TFLN photonic circuits, the intrinsic in-plane asymmetry may be exploited to improve phase-noise performance across on-chip components and routing waveguides through co-optimization of device geometry, circuit layout, optical polarization, and mode profiles, while accounting for their intended functionalities. Finally, we anticipate that further optimization of post-fabrication annealing conditions, such as temperature and cycle duration, as well as surface passivation strategies, could substantially enhance microresonator stability in ferroelectric material platforms. Overall, these results provide critical insights into the fundamental noise mechanisms in TFLN photonics and establish practical design guidelines for realizing optimized microresonators in applications demanding ultrahigh frequency stability, including optomechanical sensing, coherent optical communications, and $\chi _2 -$based squeezed-light generation.

\vspace{1em}  
\medskip
\textbf{Supporting Information} \par 
Supporting Information is available from the Wiley Online Library or from the author.

\medskip

\section{Conflict of interest}
C. L., Z.C. and M.Y. are involved in developing lithium niobate technologies at Opticore Inc..

\section{Acknowledgements}
$\,\,\,$ We thank Dr. Yun Zhao and Yuanhao liang for their insightful discussion. This work is supported by the Optica Foundation, the Chan Zuckerberg Initiative Foundation (Dynamic Imaging, 2023-321175),  the DARPA Young Faculty Award (D23AP00252-02) and DARPA under the Optomechanical Thermal Imaging (OpTIm) program (HR00112320022) and the NaPSAC program (Grant No. HR001123S0024). M.Y. and Y.Y. are supported by the U.S. Department of Energy, Office of Science, Basic Energy Sciences, Materials Sciences and Engineering Division under Contract No. DE-AC02-05CH11231 within the Quantum Coherent Systems Program KCAS26. TFLN Device fabrication was performed at the John O’Brien Nanofabrication Laboratory at University of Southern California. The SiN resonator was provided by illoomina. The views, opinions and/or findings expressed are those of the authors and should not be interpreted as representing the official views or policies of the Department of Defense or the U.S. Government.

\section{Data availability}
The data that support the findings of this study are available from the corresponding author upon reasonable request.

\section{Author contributions}
$\,\,\,$ M.Y. conceived the conceptual idea. R.Y. conducted the experiment with the help of Z.C.. Y.Y designed and fabricated the suspended chip. C.L. fabricated the cladding chip with help of R.Y. R.Y. analyzed the data with the help of I.A. R.Y. drafted the manuscript, which was revised with significant input from I.A and M.Y., and contributions from all authors. M.Y. supervised the project.

\clearpage
\bibliographystyle{unsrt}
\bibliography{references}
\medskip

\end{document}